# Twin Subsequence Search in Time Series

Georgios Chatzigeorgakidis
IMSI, Athena R.C.
gchatzi@athenarc.gr

Dimitrios Skoutas
IMSI, Athena R.C.
dskoutas@athenarc.gr

Kostas Patroumpas
IMSI, Athena R.C.
kpatro@athenarc.gr

Themis Palpanas
LIPADE, Université de Paris &
French University Institute (IUF)
themis@mi.parisdescartes.fr

Spiros Athanasiou
IMSI, Athena R.C.
spathan@athenarc.gr

Spiros Skiadopoulos
DIT, University of Peloponnese
spiros@uop.gr

## ABSTRACT

We address the problem of subsequence search in time series using Chebyshev distance, to which we refer as twin subsequence search. We first show how existing time series indices can be extended to perform twin subsequence search. Then, we introduce TS-Index, a novel index tailored to this problem. Our experimental evaluation compares these approaches against real time series datasets, and demonstrates that TS-Index can retrieve twin subsequences much faster under various query conditions. This paper has been published in the 24$^{th}$ International Conference on Extending Database Technology (EDBT 2021).

## 1 INTRODUCTION

Given a time series $T$ and a query sequence $Q$ ($|Q| \ll |T|$), *subsequence search* finds subsequences in $T$ that are similar to $Q$. Although most works rely on Euclidean distance or Dynamic Time Warping (DTW) (e.g., [17, 19]), different $\mathcal{L}_p$ norms or other similarity measures are also useful for capturing different patterns of similarity or achieving higher classification accuracy in certain datasets [6, 22]. In this work, we use the *Chebyshev* distance (i.e., $\mathcal{L}_\infty$ norm) between two subsequences, which is the maximum difference of their values across their entire duration. We call two subsequences *twins* with respect to a distance threshold $\epsilon$, if their Chebyshev distance is not greater than $\epsilon$. This kind of similarity search can be useful is various applications: finding *doublet* earthquakes in seismology, identifying similar traffic patterns in road networks, or detecting irregular patterns in medical applications like Electroencephalography (EEG) or Electrocardiography (ECG) sequences, etc.

The following indicative experiment on an EEG time series [12] with length of 1,801,999 timestamps provides some insight on the different results obtained using Chebyshev distance as opposed to Euclidean. Considering a query sequence $Q$ and a Chebyshev distance threshold $\epsilon$, we identify all twin subsequences, obtaining 1,034 results in total. We then attempt to retrieve the same results by subsequence search using Euclidean distance. To avoid any false negatives, as will be shown later in Section 3.1, we need to set the Euclidean distance threshold to $\epsilon' = \epsilon \times \sqrt{|Q|}$. The latter produces 127,887 results. Figure 1 exemplifies the intuition behind matches obtained with Chebyshev distance compared to those with Euclidean, for two different queries. Assume a query sequence $Q$ and two matches, $T$ and $T'$, obtained under Chebyshev and Euclidean distance, respectively. As shown, $T$ closely matches the query in all timestamps. Instead, $T'$ either lacks a spike that is present in the query (Fig. 1a) or exhibits one that is not present in the query (Fig. 1b).

Given a query sequence $Q$ and a time series $T$, a naïve process for finding twin subsequences of $Q$ across $T$ is by performing a sweepline scan. This scans $T$ using a sliding window of length $|Q|$, comparing at each timestamp the query with the current subsequence extracted from $T$, and adding it to the results if it satisfies the given threshold $\epsilon$ on Chebyshev distance. However, this is clearly inefficient for long time series.

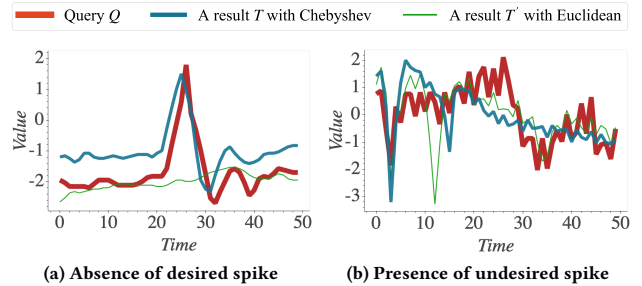

Figure 1: Examples of false positives obtained with Euclidean distance compared to results with Chebyshev distance on subsequences of the EEG dataset.

In this work, we investigate index-based methods for twin subsequence search. First, we show how two state-of-the-art time series indices, namely KV-Index [19] and *iSAX* [18] can be adapted for this task. Then, we introduce a novel index, called TS-Index, which is tailored to this problem. TS-Index is a tree structure that summarizes the subsequences contained within each node using *Minimum Bounding Time Series* (MBTS) [4], consisting of an upper and lower bounding sequence. Our experimental evaluation shows that executing twin subsequence search using TS-Index is significantly faster compared to adapting the query execution over other indices.

Specifically, our main contributions are as follows:

- We introduce the problem of twin subsequence search and propose a filter-verification algorithm that can be applied on state-of-the-art time series indices.
- We then introduce TS-Index, a tree-based index tailored to twin subsequence search, which utilizes appropriate bounds in its nodes to prune the search space.
- We experimentally evaluate our proposed methods using real-world datasets in terms of query execution, memory footprint and index construction time.

The remainder of the paper is organized as follows. Section 2 reviews related work. Section 3 formally defines the problem. Section 4 presents how it can be addressed based on existing indices. Section 5 presents the proposed TS-Index. Section 6 reports our experimental results. Finally, Section 7 concludes the paper.

## 2 RELATED WORK

Subsequence search can be performed with a sweepline approach that scans the time series using a sliding window. Various optimizations can be found in UCR suite [17] and Matrix Profile [21]. However, these optimizations are specific to Euclidean distance and thus cannot be applied to twin subsequences. Also, the lack of an index poses efficiency and scalability limitations.

A survey of time series indices for similarity search can be found in [7]. Several methods use Discrete Wavelet Transform to reduce dimensionality and then generate an index based on the transformed sequences (e.g., [3, 16]). More recent approaches are based on the *Symbolic Aggregate Approximation* (SAX) representation of time series [10]. A SAX *word* is a multi-resolution summary of a time series quantized on the value domain. It is derived from the Piecewise Aggregate Approximation (PAA) [8], which segments a time series on the time axis and approximates it by retaining only the mean value per segment. This has led to the *i*SAX index [18], a tree-based structure built over the SAX words of a set of time series. Each node in *i*SAX contains a SAX word that guarantees a lower bound in terms of Euclidean distance for all the time series indexed by it. To answer similarity search queries, the index is traversed in a top-down fashion, comparing at each step the SAX representation of the query against the ones contained in each visited node. Several extensions to *i*SAX have been proposed [13]. *i*SAX 2.0 [1] and *i*SAX2+ [2] enable bulk loading, while ADS+ [23] builds the index adaptively, based on the query workload. DP*i*SAX [20] is a distributed index. ParIS [14] and MESSI [15] take advantage of modern multi-core architectures. Coconut [9] introduces sortable SAX representations and builds an index in a bottom-up fashion. Finally, ULISSE [11] answers queries of varying length.

Another recent method for subsequence search is KV-Index [19]. After extracting all subsequences of a given length from a time series and deriving their corresponding mean values, it generates an index containing key-value pairs. Each key represents a range of mean values for a group of subsequences, pointing to starting positions of these subsequences along the original time series.

As we show in Section 4, it is possible to execute twin subsequence search queries using *i*SAX or KV-Index. However, since these indices are tailored to similarity search using Euclidean distance, this approach is suboptimal, as indicated also in our experiments in Section 6.

An index for arbitrary $\mathcal{L}_p$ norms is described in [22]. It divides each sequence into a fixed number of equi-sized segments, and takes the mean of each segment to form a feature vector. Such a generic approach favors flexibility; instead, our focus in this paper is on optimizing performance specifically for queries using Chebyshev distance.

Finally, in a previous work [5], we have studied the problem of discovering pairs and bundles of similar time-aligned subsequences within a collection of time series, based on Chebyshev distance, using a sweepline approach. In this paper, we focus on searching for twin subsequences in an input time series $T$ that are similar to a query subsequence $Q$, which is a different problem, and we propose an index-based approach. Furthermore, in another previous work [4], we have developed a hybrid index, called BTSR-Tree, which also employs the concept of Minimum Bounding Time Series (MBTS) to prune the search space. However, this is a spatial-first index specifically tailored to queries over geo-located time series, and it is based on Euclidean distance instead of Chebyshev.

## 3 PROBLEM DEFINITION

Next, we formally introduce the problem of twin subsequence search and describe a generic filter-verification approach.

### 3.1 Problem Statement

A *time series* is a time-ordered sequence $T = \{T_1, T_2, ..., T_n\}$, where $T_i$ is the value at the $i$-th timestamp and $n = |T|$ is the length of the series (i.e., number of timestamps). We use $T_{p,l}$ to denote the *subsequence* $\{T_p, ..., T_{p+l-1}\}$ starting at timestamp $p$ and having length $l$, where $1 \leq p \leq p + l - 1 \leq n$. For brevity, we also use $S$ to generally refer to a (sub)sequence.

Given two sequences $S$ and $S'$ of equal length $l$, we call them *twins* if their Chebyshev distance is not greater than a given threshold $\epsilon$. The Chebyshev distance of two vectors is their maximum difference along any dimension. Hence, if $S$ and $S'$ are twin sequences with respect to $\epsilon$, their values at any timestamp should not differ by more than $\epsilon$. Formally:

**Definition 1 (Twin Sequences).** *Two sequences $S$ and $S'$ of equal length $l$ are called* twins *with respect to a given threshold $\epsilon$, denoted as $S_1 \sim_\epsilon S_2$, if their Chebyshev distance $d$ is not greater than $\epsilon$, i.e., $d(S, S') := \max_{i=0}^{l-1}(|S_i - S'_i|) \leq \epsilon$.*

We can now formally define the problem:

**Problem 1 (Twin Subsequence Search).** *Given a query sequence $Q$ of length $l$, a time series $T$ of length $n \gg l$, and a distance threshold $\epsilon$, find all subsequences $S$ in $T$ ($|S| = l$) such that $Q \sim_\epsilon S$.*

We note two important observations below. Given two twin sequences $S \sim_\epsilon S'$ of length $l$, their Euclidean distance is $ED(S, S') = \sqrt{\sum_i (S_i - S'_i)^2} \leq \sqrt{\sum_i \epsilon^2} = \epsilon \times \sqrt{l}$. This establishes a relation between a given Chebyshev distance threshold and a corresponding Euclidean distance threshold. Moreover, from Definition 1, it follows that any pair of time-aligned subsequences across two twin sequences are also twins, i.e., if $T \sim_\epsilon T'$, then $T_{p,l} \sim_\epsilon T'_{p,l}$ for any $l \in [1, |T|]$ and $p \in [1, |T| - l]$.

Often, *z*-normalization is applied when comparing time series. Throughout the paper, we consider various possibilities: (a) working with the raw values, (b) *z*-normalizing the entire time series, (c) *z*-normalizing each individual subsequence. We discuss the implications of each case where relevant.

### 3.2 Filter-Verification Approach

We can detect twin subsequences following a filter-verification framework: the first step (*filtering*) generates candidate subsequences, which are then evaluated in the second step (*verification*) to identify those satisfying the Chebyshev distance threshold. A straightforward approach for generating candidates is to scan the entire time series $T$ with a *sweepline* and consider each subsequence $T_{p,l}$ for $p \in [1, |T| - l]$ as a candidate.

Verification is done by checking all pairwise value differences between $Q$ and $T_{p,l}$. If the difference found at a timestamp exceeds $\epsilon$, then candidate $T_{p,l}$ is rejected, otherwise it is accepted. Verification can be accelerated by detecting false positives as early as possible. If the values are *z*-normalized, we can prioritize those points in $Q$ having the highest absolute value, since these are less likely to have a match with the respective points in $T_{p,l}$. This optimization is also used in *UCR Suite* [17], and is known as *reordering early abandoning*.

The drawback of this sweepline approach is that it generates an excessive number of candidates (specifically, $|T| - l$), thus incurring a prohibitive cost when dealing with long series. To filter candidates more effectively, in the following sections we present methods based on indexing the subsequences of $T$. First, we address the problem using state-of-the-art indices; then, we introduce a novel index tailored to twin subsequence search.

## 4 TWIN SUBSEQUENCE SEARCH WITH EXISTING INDICES

Next, we focus on two representative state-of-the-art indices for time series similarity search, namely KV-Index [19] and iSAX [2], showing how they can be used for twin subsequence search without altering their structure.

### 4.1 KV-Index

Given a time series $T$, KV-Index [19] is built by considering all its subsequences of a pre-defined length $l$. Each subsequence $S$ is represented by a pair $(p, \mu)$, where $p$ is its starting position (i.e., timestamp) in $T$ and $\mu$ is its mean value over the next $l$ timestamps. KV-Index is an inverted index constructed over these pairs. Each key is a range of mean values, whereas each inverted list entry contains intervals of positions.

Twin subsequence search can be performed with KV-Index based on the following observation. If two subsequences $S$ and $S'$ of length $l$ are twins with respect to $\epsilon$, i.e., $S \sim_\epsilon S'$, then their mean values $\mu$ and $\mu'$ cannot differ by more than $\epsilon$, i.e., $|\mu - \mu'| \leq \epsilon$. Based on this, we can use a KV-Index built over a time series $T$ to generate candidates for detecting twin subsequences. Specifically, assume a query sequence $Q$ with mean value $\mu_q$. The candidate subsequences in $T$ are those included in the inverted lists with keys $[\mu_{min}, \mu_{max}]$, such that $\mu_{min} - \epsilon \leq \mu_q \leq \mu_{max} + \epsilon$. Then, the obtained candidates must be verified to derive the final results. Notice that this property is not effective if each individual subsequence has been z-normalized, because then all mean values are zero. Hence, KV-Index is applicable when working with raw values or if the entire sequence is z-normalized.

### 4.2 iSAX Index

iSAX is a tree index structure for time series similarity search [2]. Time series are z-normalized and indexed using their *Symbolic Aggregate approXimation (SAX)* [18]. The SAX representation of a series is derived in two steps. The first applies *Piecewise Aggregate Approximation* (PAA) [8], which splits the series in a specified number $m$ of segments and approximates each one with the mean value over the corresponding time interval. The second step applies quantization to assign each mean value to a discrete SAX *symbol*. Hence, each SAX symbol $X$ corresponds to a range of mean values $[\mu_{X_{min}}, \mu_{X_{max}})$. The SAX representation of a series is a sequence of $m$ SAX symbols (one symbol per segment), and is called SAX *word*. Notice that, by default, SAX words are derived using precomputed breakpoints that are selected assuming z-normalized values; nevertheless, non-normalized values can also be handled by adjusting the breakpoints accordingly.

Twin subsequence search can be enabled over iSAX by reasoning as follows. Assume two subsequences $S$ and $S'$ of length $l$, and their SAX representations $SAX(S) = \{X_1, X_2, ..., X_m\}$ and $SAX(S') = \{X'_1, X'_2, ..., X'_m\}$. As we have observed earlier, (a) if two sequences are twins with respect to a threshold $\epsilon$, then the difference between their mean values is also bounded by $\epsilon$, and (b) any pair of time-aligned segments across two twin sequences are also twins. Combining these two properties, we can see that if $S \sim_\epsilon S'$, then for each pair of symbols $X_i$ and $X'_i$ in the respective SAX representations, the mean values denoted by these symbols must not differ by more than $\epsilon$. Hence, if $S \sim_\epsilon S'$, then $\mu_{X_{i_{max}}} \geq \mu_{X'_{i_{min}}} - \epsilon$ and $\mu_{X_{i_{min}}} \leq \mu_{X'_{i_{max}}} + \epsilon$ for any $i \in [1, m]$.

Consequently, we can perform twin subsequence search using iSAX as follows. Given a time series $T$, we construct an iSAX index over all its $l$-length subsequences. Then, for a query sequence $Q$, we traverse the iSAX index starting from its root. At each node, we check the SAX word of $Q$ against the SAX word of that node, applying the property mentioned above. If the check fails, the node and its subtree can be safely pruned; otherwise, the search continues at the node's children. Once a leaf node is reached, and qualifies according to this check, all subsequences indexed therein are retrieved as candidates for verification.

## 5 THE TS-INDEX

As discussed in Section 4, it is possible to use KV-Index or iSAX to identify candidates for twin subsequence queries. However, since these indices are not tailored to the matching criterion, they tend to generate a large number of false positives, incurring a significant verification cost, as confirmed in our experiments. In the following, we introduce TS-Index, which is specifically designed for twin subsequence search. First, we provide an overview of its structure and explain how it is constructed. Then, we present an algorithm to evaluate twin subsequence queries specifying a distance threshold.

### 5.1 Index Structure

The core concept in TS-Index is that of *Minimum Bounding Time Series* (MBTS) [4]. An MBTS is a pair of sequences that fully encloses a set of time series $\mathcal{T}$ by indicating the maximum and minimum values at each timestamp. Figure 2a depicts an example of an MBTS enclosing a set of four time series. Formally:

DEFINITION 2 (MBTS). *Given a set $\mathcal{T}$ of time series with equal length $l$, its MBTS $B = (B^\sqcap, B^\sqcup)$ consists of an upper bounding time series $B^\sqcap$ and a lower bounding time series $B^\sqcup$, constructed by respectively selecting the maximum and minimum values at each timestamp $i \in \{1, \ldots, l\}$ among all time series in $\mathcal{T}$ as follows:*

$$B^\sqcap = \{\max_{T \in \mathcal{T}} T_1, \ldots, \max_{T \in \mathcal{T}} T_l\}$$
$$B^\sqcup = \{\min_{T \in \mathcal{T}} T_1, \ldots, \min_{T \in \mathcal{T}} T_l\} \quad (1)$$

The TS-Index has a tree structure. Each internal node points to a set of children nodes, whereas each leaf node points to a set of subsequences (more specifically, to the starting positions of its indexed subsequences along the input time series $T$). All leaf nodes are at the same level. Each node is associated with an MBTS, which encloses all the sequences indexed therein. Clearly, MBTS get tighter when descending from the root to the leaf level. Figure 3a illustrates an example of TS-Index for nine input sequences. The MBTS of each node is depicted as a grey band.

### 5.2 Index Construction

Assume an input time series $T$ and a subsequence length $l$. The TS-Index over $T$ is constructed in a top-down fashion, by sequentially inserting all $l$-length subsequences of $T$. When inserting a sequence $S$, we traverse the index from the root, selecting at each level the node whose MBTS has the smallest distance from $S$, until a leaf node is reached. The distance between a sequence $S$ and an MBTS $B$ is calculated using the following formula:

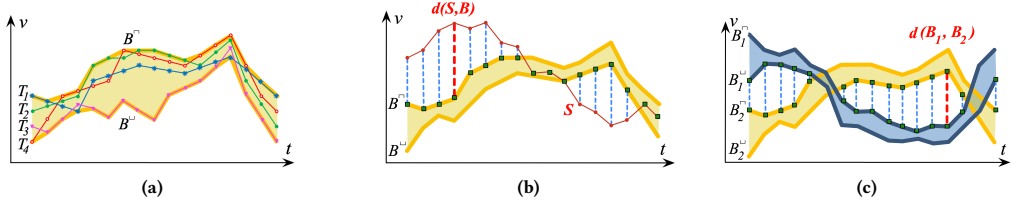

Figure 2: (a) MBTS enclosing a set of 4 time series. Distance between (b) a sequence $S$ and an MBTS $B$, (c) MBTS $B_1$ and $B_2$.

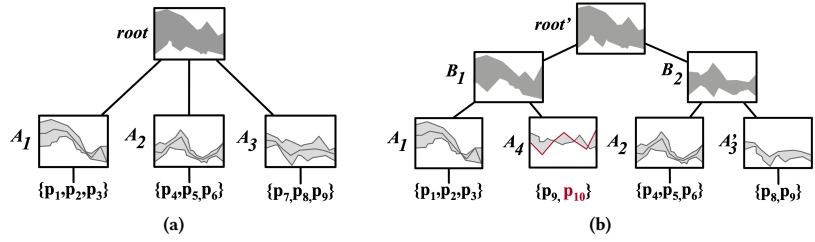

Figure 3: (a) TS-Index for 9 input sequences. (b) Inserting $p_{10}$ causes a split at leaf $A_3$ and splits propagate upwards.

$$d(S, B) = \max_i \begin{cases} S_i - B_i^\sqcap & \text{if } S_i > B_i^\sqcap \\ B_i^\sqcup - S_i & \text{if } S_i < B_i^\sqcup \\ 0 & \text{otherwise} \end{cases} \quad (2)$$

where $B_i^\sqcap$ and $B_i^\sqcup$ are the $i^{th}$ values of the upper and lower bounds of the MBTS $B$, respectively.

Each node has a minimum capacity $\mu_c$ and a maximum *capacity* $M_c$, specifying the minimum and maximum number of children it can point to. Once a node exceeds $M_c$, it is split in two nodes. This may cause the parent node to also exceed the maximum capacity $M_c$, in which case it is split too. Hence, this process recursively propagates upwards until no further splits occur. This procedure ensures that all leaves are placed on the same level of the tree.

During node splitting, the goal is to make the MBTS of each new sibling node as tight as possible. If this is a leaf node, we identify the two subsequences within the original node having the highest Chebyshev distance and use them as seeds for the two sibling nodes. Each remaining subsequence is assigned to the node where it causes the smallest expansion of its MBTS, which gets updated accordingly. For an internal node, the process is similar. Yet, adjusting its MBTS in this case involves the MBTS of children nodes instead of individual sequences. To accommodate this, the distance between two MBTS $B_1$ and $B_2$ is defined as:

$$d(B_1, B_2) = \max_i \begin{cases} B_{1,i}^\sqcup - B_{2,i}^\sqcap & \text{if } B_{1,i}^\sqcup > B_{2,i}^\sqcap \\ B_{2,i}^\sqcup - B_{1,i}^\sqcap & \text{if } B_{1,i}^\sqcap < B_{2,i}^\sqcup \\ 0 & \text{otherwise} \end{cases} \quad (3)$$

where $B_{1,i}^\sqcup, B_{1,i}^\sqcap$ and $B_{2,i}^\sqcup, B_{2,i}^\sqcap$ are the $i^{th}$ values of the upper and lower bounds of the MBTS $B_1$ and $B_2$, respectively. Figures 2b and 2c exemplify the calculation of the distance of a sequence $S$ to an MBTS $B$ and the calculation of the distance between two MBTS ($B_1, B_2$) respectively; in both cases, the distance is the length of the dashed red line.

Figure 3b depicts an example where inserting subsequence $p_{10}$ into leaf node $A_3$ of the TS-Index in Figure 3a, causes it to split into two new nodes, $A_3'$ and $A_4$ (we assume $\mu_c = 2$ and $M_c=3$). This process is then propagated upwards, splitting the root into $B_1$ and $B_2$. To keep the MBTS tight –according to Equation 3–, nodes $A_1$, $A_4$ have become children of $B_1$ and $A_2$, $A_3'$ are now children of $B_2$. Finally, a new root is added, increasing the index height by one.

### 5.3 Query Execution

Twin subsequence search can be performed with TS-Index based on the following lemma.

LEMMA 1. *Assume a query sequence $Q$ and a node $N$ of the TS-Index with MBTS $B$. If there exists a sequence $S$ indexed at $N$ such that $Q \sim_\epsilon S$, then $d(Q, B) \leq \epsilon$.*

PROOF. Assume that $Q \sim_\epsilon S$ for a sequence $S$ indexed by node $N$. From Definition 2, it follows that $S_i \in [B_i^\sqcup, B_i^\sqcap]$ for each timestamp $i$. Moreover, from Definition 1, it follows that $|Q_i - S_i| \leq \epsilon$. Hence, from Equation 2, we derive $d(Q, B) \leq \epsilon$. □

Given a query sequence $Q$, we traverse the index in a top-down fashion, starting from its root. For each visited node $N$, we compare $Q$ against $N$'s MBTS, applying Lemma 1 to prune its subtree. Note that this check can be accelerated, since it is not necessary to fully compute distance $d(Q, B)$; instead, if the indexed values have been $z$-normalized, we apply early abandoning (see Section 3.2) to prune the node as soon as the value difference exceeds $\epsilon$ in at least one timestamp. Multiple paths starting from the root may need to be explored, depending on the query and the tightness of the bounds in the visited nodes.

Algorithm 1 describes the search process. The input includes the query sequence $Q$, the constructed TS-Index $I$, the given time series $T$ and the threshold $\epsilon$. We start by initializing a list $L$ with the root's children (Line 2). Then, we traverse the index by iterating over this list (Lines 3-12). For each node $N$ currently in the list, we obtain its MBTS (Lines 4-5). Then, we check whether the distance between this MBTS and the query is higher than the specified threshold $\epsilon$ (Line 6). If so, the subtree under the current node $N$ is pruned; otherwise, it is examined as explained next. If $N$ is not a leaf node, we insert its children in list $L$ for probing (Lines 7-8). Once a leaf node is reached, we iterate over all the subsequence positions it contains and check whether each corresponding subsequence is a twin of $Q$ with respect to $\epsilon$. If so, we add this subsequence to the final results (Lines 9-12). The results are returned once all candidate nodes in list $L$ have been either probed or pruned (Line 13).

**Algorithm 1:** TwinSubsequenceSearch

**Input** : Time series $T$, TS-Index $I$, query $Q$, threshold $\epsilon$
**Output:** List $R$ of twin subsequences to $Q$

1  $R \leftarrow \emptyset$
2  $L \leftarrow I.root.\text{getChildren}()$
3  **while** $L \neq \emptyset$ **do**
4     $N \leftarrow L.\text{getNext}()$
5     $B \leftarrow N.MBTS$
6     **if** $d((Q, B) \leq \epsilon$ **then**
7       **if** $N$ *is not leaf* **then**
8         $L \leftarrow L \cup \{N.\text{getChildren}()\}$
9       **else**
10        **foreach** $p \in N.\text{getPositions}()$ **do**
11          **if** $d((Q, T_{p,l}) \leq \epsilon$ **then**
12            $R \leftarrow R \cup T_{p,l}$

13 **return** R

**Table 1: Datasets and distance thresholds.**

| Dataset | $n$ | $\epsilon$ (norm) | $\epsilon$ (non-norm) |
|---|---|---|---|
| Insect | 64,436 | 0.5,**0.75**,1,1.25,1.5 | 50, **100**,150,200,250 |
| EEG | 1,801,999 | 0.1,0.2,**0.3**,0.4,0.5 | 20, **40**,60,80,100 |

**Table 2: Other parameters.**

| Parameter | Value |
|---|---|
| Number $m$ of segments | 5, **10**, 20, 25, 50 |
| Sequence length $l$ | 50, **100**, 150, 200, 250 |

## 6 EXPERIMENTAL EVALUATION

Next, we present an experimental evaluation of our methods against two real-world datasets.

### 6.1 Experimental Setup

We performed experiments against two real-world time series (see Table 1), which contain diverse patterns and differ in their total duration. In particular, the *Insect Movement* [12] series contains 64,436 insect telemetry readings spanning around 30 minutes (36 readings/sec), whereas the *Electroencephalography (EEG)* [12] series comprises 1,801,999 EEG readings at 500Hz lasting one hour. Unless stated otherwise, we $z$-normalize the time series to facilitate selection of distance thresholds.

Table 1 indicates the different values for the distance threshold $\epsilon$ used in the experiments against each dataset, for $z$-normalized (*norm*) or original values (*non-norm*). Table 2 contains the values for subsequence length $l$ and number of segments $m$, which are common in the experiments on both datasets. In both tables, default values are in bold. These values have been selected after running several preliminary tests, which also guided selection of other parameters. Specifically, for *i*SAX, the maximum node capacity is set to 10,000 to enable index construction in reasonable time even for larger datasets. The default values for minimum and maximum node capacity in TS-Index are set to $\mu_c = 10$ and $M_c = 30$, respectively.

For each dataset, we randomly picked 100 subsequences, each of length $l = 100$ points, and used them as the query workload in all tests against that dataset. We report average response time per query (in milliseconds). We implemented all methods, including KV-Index, *i*SAX, and TS-Index, in Java. In all implementations, the structure of the index is kept in memory, while the original input dataset is stored on disk. Leaf nodes in the index contain the starting positions of the subsequences in the input time series. Thus, when a leaf is reached at query time, its corresponding subsequences are obtained from the input time series file using random access. All experiments were conducted on a server with

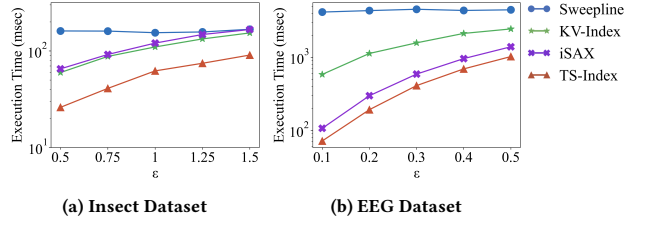

(a) Insect Dataset      (b) EEG Dataset

**Figure 4: Varying distance threshold $\epsilon$.**

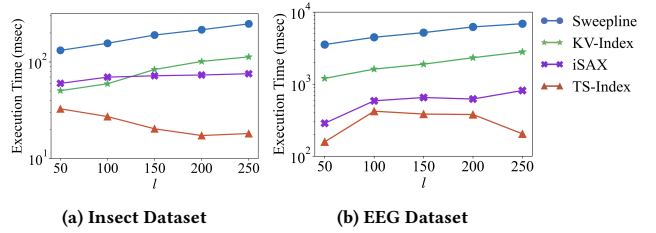

(a) Insect Dataset      (b) EEG Dataset

**Figure 5: Varying subsequence length $l$.**

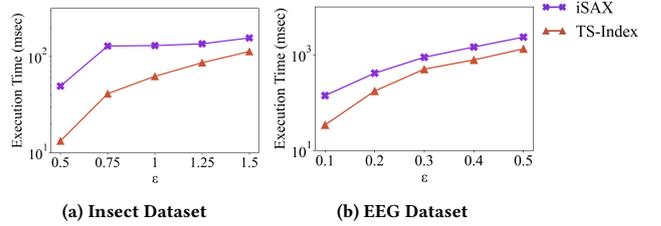

(a) Insect Dataset      (b) EEG Dataset

**Figure 6: Varying $\epsilon$ on $z$-normalized subsequences.**

4 CPUs, each equipped with 8 cores clocked at 2.13GHz, and 256 GB RAM running Debian Linux.

### 6.2 Performance

We compare the average execution time per query for varying values of each parameter, setting the rest to their default values.

*6.2.1 Varying threshold $\epsilon$.* Figure 4 depicts query execution time (in logarithmic scale) for varying threshold $\epsilon$. As expected, searching with the Sweepline approach has a fixed cost per dataset regardless of $\epsilon$, since it needs to scan all subsequences extracted from the input time series. Relaxing the threshold incurs an overhead when an index is involved. Queries against KV-Index perform poorly compared to other indices, since filtering based on mean values achieves less pruning. Searching with TS-Index outperforms the rest in every setting for both tested datasets. Overall, TS-Index is at least an order of magnitude more efficient in twin subsequence search compared to the KV-Index and Sweepline approaches. It is also consistently better than *i*SAX as it is less susceptible to fluctuations in the input sequences.

*6.2.2 Varying Subsequence Length.* Figure 5 plots performance results with a varying length $l$ for subsequences obtained from the input time series. Increasing $l$ seems to slightly negatively affect all approaches, except for TS-Index. Since longer subsequences are extracted, more checks are required, both in nodes (in case of *i*SAX) and raw subsequences during verification. Instead, TS-Index is faster when longer subsequences are specified, as it becomes less likely to find matching twins. In particular, TS-Index has higher pruning capability and can skip non-qualifying subtrees earlier at higher levels in the tree hierarchy. Thus, fewer leaf nodes are accessed and need to be verified, saving much of the verification cost for checks per timestamp.

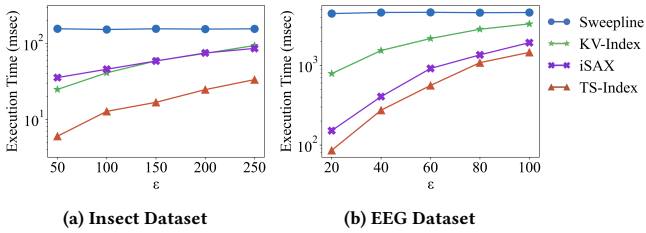

(a) Insect Dataset  (b) EEG Dataset

**Figure 7: Varying $\epsilon$ on non-normalized data.**

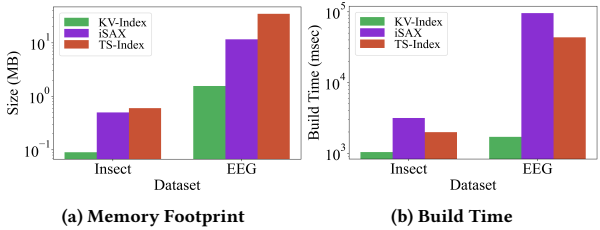

(a) Memory Footprint  (b) Build Time

**Figure 8: Memory footprint and build time per index.**

*6.2.3 Searching over z-normalized subsequences.* We repeat the experiment for varying distance threshold $\epsilon$, this time applying z-normalization over each individual subsequence, before inserting it in the index. As mentioned in Section 4.1, KV-Index cannot be built on such data since the mean value per subsequence would always be zero; thus, we only compare TS-Index with *i*SAX. The results are depicted in Figure 6. Clearly, z-normalizing the subsequences separately has no significant effect on the performance of TS-Index; the results are similar to those in Figure 4, with TS-Index outperforming *i*SAX in all cases.

*6.2.4 Searching on Non-Normalized Data.* Query execution cost for identifying twin subsequences against the raw (non-normalized) time series is depicted in Figure 7. Overall, TS-Index copes better than all the rest even for raw data, confirming its suitability for twin subsequence search in various settings.

*6.2.5 Index Size.* Figure 8a presents the memory footprint of TS-Index, *i*SAX and KV-Index for each dataset. KV-Index requires less space than TS-Index and *i*SAX, as it only keeps in memory the mean value and position range per subsequence. Instead, TS-Index and *i*SAX occupy more space due to their more complex structures. Specifically, *i*SAX requires two to three times less space than TS-Index. Indeed, *i*SAX needs to store one SAX word per node, whereas a node in TS-Index is represented by an MBTS, hence its increased memory footprint. Nevertheless, all indices, including TS-Index, have sizes that easily fit in main memory.

*6.2.6 Build Time.* Similarly to the index size, and due to the significantly less required calculations (i.e., only subsequence mean values need be calculated and no node splitting is needed), KV-Index requires significantly less time to be constructed than *i*SAX and TS-Index (Figure 8b). *i*SAX is the slowest index to be built, since it needs to additionally convert the PAA of each subsequence to a SAX word for each extracted subsequence.

## 7 CONCLUSIONS

In this paper, we have introduced the twin subsequence search problem. Given a query sequence $Q$, an input time series $T$ and a distance threshold $\epsilon$, this task retrieves all subsequences in $T$ with Chebyshev distance to $Q$ not higher than $\epsilon$. To answer this query efficiently, we have introduced the TS-Index. We have described the index structure and proposed algorithms for efficient index construction and query answering. Our experimental evaluation assesses the TS-Index in terms of construction cost and confirms its superiority for twin subsequence search queries when compared to the state-of-the-art.